\documentstyle[12pt]{article}
\newlength{\pubnumber} 

\catcode`\@=11
\@addtoreset{equation}{section}
\def\section{\@startsection{section}{1}{\z@}{3.5ex plus 1ex minus .2ex}
 {2.3ex plus .2ex}{\large\bf}}
\def\subsection{\@startsection{subsection}{2}{\z@}{2.3ex plus .2ex}
 {2.3ex plus .2ex}{\bf}}

    \renewcommand{\baselinestretch}{1.4}
\begin{document}

\begin{titlepage}
\samepage{
\setcounter{page}{1}
\rightline{UFIFT-HEP-96-11}
\rightline{\tt hep-ph/9604302}
\rightline{April 1996}
\vfill
\begin{center}
 {\Large \bf  Leptophobic $Z^\prime$ \\
     From Superstring Derived Models\\}
\vfill
 {\large Alon E. Faraggi$^1$\footnote{
   E-mail address: faraggi@phys.ufl.edu}
   $\,$and$\,$ Manuel Masip$^{1,2}$\footnote{
   E-mail address: masip@ugr.es}\\}
\vspace{.12in}
 {\it  $^{1}$ Institute For Fundamental Theory, \\
  University of Florida, 
  Gainesville, FL 32611 USA\\}
\vspace{.05in}
 {\it  $^{2}$ Departamento de F\'\i sica Te\'orica y del Cosmos, \\
		Universidad de Granada,
		18071 Granada, Spain\\}
\end{center}
\vfill
\begin{abstract}
  {\rm
It was recently suggested that the reported anomalies in $R_b$ 
and $R_c$ can be interpreted as the effect of a heavy vector 
boson that couples to quarks and is universally decoupled from leptons. 
We examine how an extra gauge boson with this property 
can arise from superstring derived models. 
In a specific three generation model we show 
that the $U(1)_{B-L}$ symmetry combines with the 
horizontal flavor symmetries to form a universal leptophobic 
$U(1)$ symmetry. 
In our model there 
is an enhancement of the color gauge group from twisted sectors. 
The enhancement occurs after the breaking of the 
unifying gauge symmetry by ``Wilson lines''.
The leptophobic $U(1)$ symmetry then becomes a generator of the 
color $SU(4)$ gauge group. We examine how similar symmetries may 
appear in other string models without the enhancement. 
We propose that if the current LEP anomalies 
persist it may be evidence for a certain class of 
un--unified superstring models. 

}
\end{abstract}
\vfill
\smallskip}
\end{titlepage}

\setcounter{footnote}{0}

\def\beq{\begin{equation}}
\def\eeq{\end{equation}}
\def\beqn{\begin{eqnarray}}
\def\eeqn{\end{eqnarray}}
\def\AEF{A.E. Faraggi}
\def\NPB#1#2#3{{\it Nucl.\ Phys.}\/ {\bf B#1} (19#2) #3}
\def\PLB#1#2#3{{\it Phys.\ Lett.}\/ {\bf B#1} (19#2) #3}
\def\PRD#1#2#3{{\it Phys.\ Rev.}\/ {\bf D#1} (19#2) #3}
\def\PRL#1#2#3{{\it Phys.\ Rev.\ Lett.}\/ {\bf #1} (19#2) #3}
\def\PRT#1#2#3{{\it Phys.\ Rep.}\/ {\bf#1} (19#2) #3}
\def\MODA#1#2#3{{\it Mod.\ Phys.\ Lett.}\/ {\bf A#1} (19#2) #3}
\def\IJMP#1#2#3{{\it Int.\ J.\ Mod.\ Phys.}\/ {\bf A#1} (19#2) #3}
\def\nuvc#1#2#3{{\it Nuovo Cimento}\/ {\bf #1A} (#2) #3}
\def\etal{{\it et al,\/}\ }
\hyphenation{su-per-sym-met-ric non-su-per-sym-met-ric}
\hyphenation{space-time-super-sym-met-ric}
\hyphenation{mod-u-lar mod-u-lar--in-var-i-ant}


\setcounter{footnote}{0}

Over the past few years LEP, SLC and the Tevatron experiments provided 
impressive confirmation of the Standard Model of particle physics and its 
gauge symmetry structure. 
Recently, however, there has been accumulating evidence 
at LEP that indicates deviation from the Standard Model predictions 
in the hadronic partial width at the $Z$--boson peak \cite{lep}, which 
is commonly referred to as the $R_b-R_c$ crisis. 

It was recently 
suggested by several groups \cite{evb} that the discrepancy between the 
predicted and measured values for the decay of the $Z$--boson to 
$b$ and $c$--quarks could be explained by an additional heavy
gauge boson that couples to quarks but is universally decoupled from 
leptons. 
If this interpretation of the data is correct it will
have profound implication on attempts to understand 
the origin of the gauge and matter structure of the 
Standard Model. It might for example invalidate
the traditional approaches to embed the Standard 
Model in a simple Grand Unified Group 
as those in their nature unify the interactions 
between quarks and leptons. 

In superstring models one also traditionally 
starts with an underlying unifying gauge group
which is then broken to the Standard Model 
by using string and field theoretic 
symmetry breaking mechanisms. However, as the 
rank of the gauge group in string models is 
larger than those which are  
used in Grand Unified Theories, one can contemplate 
the possibility that a particular combination of the 
additional $U(1)$ generators in the four
dimensional Cartan subalgebra will combine to 
form a leptophobic $U(1)$. Moreover, in a generic level one 
string model, massless states that produce the Standard Model representations, 
in general, must be charged with respect to additional
$U(1)$ or discrete symmetries. As the 
assignment of charges to the quarks and lepton depend 
on the specific compactification, it is difficult 
to envision how a generic compactification will produce
the universal charge assignment which is needed. 
Furthermore, the charges of the quarks and leptons 
under the additional $U(1)$ symmetries of the four 
dimensional gauge group depend on specific 
patterns of symmetry breaking in the string models. 
Therefore, if such a leptophobic $U(1)$ symmetry is produced 
it might be a peculiar accident of a particular 
string model or perhaps of a class of string models. 

In this paper, we examine how a leptophobic $U(1)$ can arise 
from superstring derived models. The leptophobia of the $U(1)$ symmetry
is obtained by combining the $U(1)_{B-L}$ generator
with a combination of the flavor $U(1)$ symmetries. 
We present an explicit superstring 
derived model which gives rise to a universal 
leptophobic $U(1)$ symmetry. 
The leptophobic $U(1)$ symmetry is obtained in a class of 
superstring derived standard--like models \cite{SLM} 
due to a combination of 
the $U(1)_{B-L}$ symmetry, which is embedded in $SO(10)$, plus a 
combination of additional $U(1)$ symmetries. 
These additional $U(1)$ symmetries 
compensate the lepton number in $U(1)_{B-L}$
and the resulting $U(1)$ therefore becomes a gauged baryon number.
In the specific model that we study in some detail the gauge 
symmetry is enhanced due to gauge bosons from twisted 
sectors. The color $SU(3)$ gauge group is enhanced to 
$SU(4)$, and $U(1)_B$ is the $U(1)$ in the decomposition 
$SU(4)_C\rightarrow SU(3)_C\times U(1)_B$. 
Due to a symmetry between 
the three chiral generation, 
we argue that leptophobic $U(1)$ symmetries 
may in fact be common in this class of 
superstring compactification,
without further enhancement of the gauge group. 

The superstring models that we discuss are constructed in the 
free fermionic formulation \cite{FFF}. 
In this formulation a model is constructed by 
choosing a consistent set of boundary condition basis vectors.
The basis vectors, $b_k$, span a finite  
additive group $\Xi=\sum_k{{n_k}{b_k}}$
where $n_k=0,\cdots,{{N_{z_k}}-1}$.
The physical massless states in the Hilbert space of a given sector
$\alpha\in{\Xi}$, are obtained by acting on the vacuum with 
bosonic and fermionic operators and by
applying the generalized GSO projections. The $U(1)$
charges, $Q(f)$, with respect to the unbroken Cartan generators of the four 
dimensional gauge group, which are in one 
to one correspondence with the $U(1)$
currents ${f^*}f$ for each complex fermion f, are given by:
\begin{equation}
{Q(f) = {1\over 2}\alpha(f) + F(f)},
\label{u1charges}
\end{equation}
where $\alpha(f)$ is the boundary condition of the world--sheet fermion $f$
 in the sector $\alpha$, and 
$F_\alpha(f)$ is a fermion number operator counting each mode of 
$f$ once (and if $f$ is complex, $f^*$ minus once). 
For periodic fermions,
$\alpha(f)=1$, the vacuum is a spinor in order to represent the Clifford
algebra of the corresponding zero modes. 
For each periodic complex fermion $f$
there are two degenerate vacua ${\vert +\rangle},{\vert -\rangle}$ , 
annihilated by the zero modes $f_0$ and
${{f_0}^*}$ and with fermion numbers  $F(f)=0,-1$, respectively. 

The realistic models in the free fermionic formulation are generated by 
a basis of boundary condition vectors for all world--sheet fermions 
\cite{REVAMP,SLM,FNY,alr,EU,PRICE,custodial}. 
The basis is constructed in two stages. The first stage consist
of the NAHE set \cite{REVAMP,SLM}, 
which is a set of five boundary condition basis 
vectors, $\{{{\bf 1},S,b_1,b_2,b_3}\}$. The gauge group after the NAHE set
is $SO(10)\times SO(6)^3\times E_8$ with $N=1$ space--time supersymmetry. 
The vector $S$ is the supersymmetry generator and the superpartners of
the states from a given sector $\alpha$ are obtained from the sector
$S+\alpha$. The space--time vector bosons that generate the gauge group 
arise from the Neveu--Schwarz sector and from the sector $1+b_1+b_2+b_3$. 
The Neveu--Schwarz sector produces the generators of 
$SO(10)\times SO(6)^3\times SO(16)$. The sector $1+b_1+b_2+b_3$
produces the spinorial 128 of $SO(16)$ and completes the hidden 
gauge group to $E_8$. The vectors $b_1$, $b_2$ and $b_3$ 
produce 48 spinorial 16 of $SO(10)$, sixteen from each sector $b_1$, 
$b_2$ and $b_3$. The vacuum of these sectors contains eight periodic 
fermions. Five of those periodic fermions produce the charges under the 
$SO(10)$ group, while the remaining three periodic fermions 
generate charges with respect to the flavor symmetries. Each of the 
sectors $b_1$, $b_2$ and $b_3$ is charged with respect to a different set 
of flavor quantum numbers, $SO(6)_{1,2,3}$.

The NAHE set divides the 44 right--moving and 20 left--moving real internal
fermions in the following way: ${\bar\psi}^{1,\cdots,5}$ are complex and
produce the observable $SO(10)$ symmetry; ${\bar\phi}^{1,\cdots,8}$ are
complex and produce the hidden $E_8$ gauge group;
$\{{\bar\eta}^1,{\bar y}^{3,\cdots,6}\}$, $\{{\bar\eta}^2,{\bar y}^{1,2}
,{\bar\omega}^{5,6}\}$, $\{{\bar\eta}^3,{\bar\omega}^{1,\cdots,4}\}$
give rise to the three horizontal $SO(6)$ symmetries. The left--moving
$\{y,\omega\}$ states are divided into, $\{{y}^{3,\cdots,6}\}$, $\{{y}^{1,2}
,{\omega}^{5,6}\}$, $\{{\omega}^{1,\cdots,4}\}$. The left--moving
$\chi^{12},\chi^{34},\chi^{56}$ states carry the supersymmetry charges.
Each sector $b_1$, $b_2$ and $b_3$ carries periodic boundary conditions
under $(\psi^\mu\vert{\bar\psi}^{1,\cdots,5})$ and one of the three groups:
$(\chi_{12},\{y^{3,\cdots,6}\vert{\bar y}^{3,\cdots6}\},{\bar\eta}^1)$,
$(\chi_{34},\{y^{1,2},\omega^{5,6}\vert{\bar y}^{1,2}{\bar\omega}^{5,6}\},
{\bar\eta}^2)$, 
$(\chi_{56},\{\omega^{1,\cdots,4}\vert{\bar\omega}^{1,
\cdots4}\},{\bar\eta}^3)$. 

The division of the internal fermions is a
reflection of the underlying $Z_2\times Z_2$ orbifold 
compactification \cite{FOC}. 
The Neveu--Schwarz sector corresponds to the
untwisted sector, and the sectors $b_1$, $b_2$ and $b_3$ correspond to the
three twisted sectors of the $Z_2\times Z_2$ orbifold models.
At this level there is a discrete $S_3$ 
permutation symmetry between the three sectors $b_1$, $b_2$ 
and $b_3$. This permutation symmetry arises due to the symmetry of the 
NAHE set and may be essential for the universality of the leptophobic
$U(1)$ symmetry. Because of the underlying $Z_2\times Z_2$ orbifold
compactification, each of the chiral generations from the sectors 
$b_1$, $b_2$ and $b_3$ is charged with respect to a different set 
of flavor charges. 

The second stage of the basis construction consist of adding three
additional basis vectors to the NAHE set. The three additional basis 
vectors correspond to ``Wilson lines'' in the orbifold formulation. 
Three additional vectors are needed to reduce the number of generations 
to three, one from each sector $b_1$, $b_2$ and $b_3$. 
One specific example is given in table 1. 
The choice of boundary
conditions to the set of real internal fermions
${\{y,\omega\vert{\bar y},{\bar\omega}\}^{1,\cdots,6}}$  
determines the low energy properties, like the number of generations,
Higgs doublet--triplet splitting and Yukawa couplings. 

The final gauge group arises as follows. 
The NS sector produces the 
generators of $SU(3)_C\times SU(2)_L\times U(1)_C\times U(1)_L
\times U(1)_{1,2,3}\times U(1)_{4,5,6}\times SU(3)\times SO(4)\times 
U(1)_H\times U(1)_{7,8,9}$.  
The $SO(10)$ symmetry is broken to 
$SU(3)_C\times U(1)_C\times SU(2)_L\times U(1)_L$\footnote{
      			 $U(1)_C={3\over2}U(1)_{B-L}$ and
      			 $U(1)_L=2U(1)_{T_{3_R}}.$}, where
\begin{eqnarray}
U(1)_C&&={\rm Tr} U(3)_C~\Rightarrow~Q_C=
			\sum_{i=1}^3Q({\bar\psi}^i), \nonumber\\
U(1)_L&&={\rm Tr} U(2)_L~\Rightarrow~Q_L=
			\sum_{i=4}^5Q({\bar\psi}^i).
\label{u1cl}
\end{eqnarray}
The flavor $SO(6)^3$ symmetries are broken to $U(1)^{3+n}$ with
$(n=0,\cdots,6)$. The first three, denoted by $U(1)_{r_j}$, arise 
from the world--sheet currents ${\bar\eta}^j{\bar\eta}^{j^*}$
$(j=1,2,3)$. These three $U(1)$ symmetries are present in all
the three generation free fermionic models which use the NAHE set. 
Additional horizontal $U(1)$ symmetries, denoted by $U(1)_{r_j}$ 
$(j=4,5,...)$, arise by pairing two real fermions from the sets
$\{{\bar y}^{3,\cdots,6}\}$, 
$\{{\bar y}^{1,2},{\bar\omega}^{5,6}\}$, and
$\{{\bar\omega}^{1,\cdots,4}\}$. 
The final observable gauge group depends on
the number of such pairings. In the model of table 1 there are 
three such pairings, ${\bar y}^3{\bar y}^6$, ${\bar y}^1{\bar\omega}^5$
and ${\bar\omega}^2{\bar\omega}^4$, which generate three additional 
$U(1)$ symmetries, denoted by $U(1)_{r_{4,5,6}}$. It is 
important to note that the existence of these three additional 
$U(1)$ currents is correlated with a superstringy doublet--triplet
splitting mechanism \cite{ps}. Due to these extra $U(1)$ symmetries 
the color triplets from the NS sector are projected out of the spectrum 
by the GSO projections while the electroweak doublets remain in the 
light spectrum. 
The remaining $U(1)$ generators are 
\begin{eqnarray}
U(1)_H&&= {\rm Tr} U(3)_H~\Rightarrow~Q_H=\sum_{i=5}^7Q({\bar\phi}^i).
\label{u1h}
\end{eqnarray}
and $U(1)_{7,8,9}$, which arise from the world--sheet currents 
${\bar\phi}^1{\bar\phi}^{1^*}$, ${\bar\phi}^2{\bar\phi}^{2^*}$, 
${\bar\phi}^8{\bar\phi}^{8^*}$, 
respectively. 
The sector $1+b_1+b_2+b_3$ produces the representations 
$(3,2)_{-5}\oplus({\bar3},2)_5$ and $2_{-3}\oplus2_3$ of 
$SU(3)\times SU(2)_r\times U(1)_{h5}$ and $SU(2)_\ell\times U(1)_{h3}$ 
respectively, where 
$SU(2)_r\times SU(2)_\ell$ are the two $SU(2)$'s in the isomorphism
$SO(4)\sim SU(2)_r\times SU(2)_\ell$. 
Thus, the $E_8$ symmetry
reduces to $SU(5)\times SU(3)\times U(1)^2$. 
The $U(1)$'s in $SU(5)$ and $SU(3)$ are given by 
$U(1)_{h5}=-3U_7+3U_8+U_H-3U_9$ and $U(1)_{h3}=U_7+U_8+U_H+U_9$
respectively.  
The remaining $U(1)$ symmetries in the
hidden sector, $U(1)_{7^\prime}$ and $U(1)_{8^\prime}$,
correspond to the world--sheet currents
${\bar\phi}^1{\bar\phi}^{1^*}-{\bar\phi}^8{\bar\phi}^{8^*}$ and
$-2{\bar\phi}^j{\bar\phi}^{j^*}+{\bar\phi}^1{\bar\phi}^{1^*}
+4{\bar\phi}^2{\bar\phi}^{2^*}+{\bar\phi}^8{\bar\phi}^{8^*}$ respectively,
where summation on $j=5,\cdots,7$ is implied.

For some choices of the additional basis vectors that extend the NAHE 
set, there may exist a combination 
\begin{equation}
X=n_\alpha\alpha+n_\beta\beta+n_\gamma\gamma\label{xcomb}
\end{equation}
for which $X_L\cdot X_L=0$ and $X_R\cdot X_R\ne0$. Such a 
combination may produce additional space--time vector 
bosons, depending on the choice of GSO phases. 
In the model of table 1
additional space--time vector bosons are obtained from the sector 
$1+\alpha+2\gamma$ \cite{custodial}. 
The model of table 1 differs from the model of Ref. \cite{custodial} 
by a change of a GSO phase 
\begin{equation}
c\left(\matrix{\gamma\cr1\cr}\right)=-1\rightarrow
c\left(\matrix{\gamma\cr1\cr}\right)=+1
\end{equation}

In the model of table 1,
the sector $1+\alpha+2\gamma$ produces six additional space--time 
vector bosons, which are triplets of $SU(3)_C$ 
and carry $U(1)$ charges. 
One combination of the $U(1)$ symmetries
\begin{equation}
U(1)_B={1\over3}U_C-(U_{r_4}+U_{r_5}+U_{r_6})-U_{7^\prime}
\label{u1b}
\end{equation}
is the $U(1)$ generator of the enhanced color $SU(4)$ symmetry. 
The six space--time vector bosons from the sector 
$1+\alpha+2\gamma$ complete the adjoint representation of the 
gauge group. The remaining orthogonal $U(1)$ combinations are 
\begin{eqnarray}
U_{C^\prime}&&=U_C+{1\over2}U_{7^\prime},\nonumber\\
U_{4^\prime}&&=U_4-U_5,\nonumber\\
U_{5^\prime}&&=U_4+U_5-2U_6,\nonumber\\
U_{7^{\prime\prime}}&&=U_C+{7\over3}(U_4+U_5+U_6)-3U_{7^\prime}.
\label{u1com}
\end{eqnarray}
The full massless spectrum now transforms under the final gauge group, 
$SU(4)_C\times SU(2)_L\times U(1)_{C^\prime}\times U(1)_L\times
U(1)_{1,2,3}\times U(1)_{4^\prime}\times U(1)_{5^\prime}\times 
U(1)_{7^{\prime\prime}}\times U(1)_8$. 
The weak hypercharge is given by 
$U(1)_Y=1/3U(1)_{C^\prime}+1/2U(1)_L$. 
The Neveu--Schwarz sector gives, in addition to the graviton, 
dilaton, antisymmetric sector and spin 1 gauge bosons, three pairs 
of electroweak doublets, three pairs of $SO(10)$ singlets with 
$U(1)_{1,2,3}$ charges and three singlets of the entire four 
dimensional gauge group. The sector $S+b_1+b_2+\alpha+\beta$ 
produces two pairs of electroweak doublets and four pairs 
of $SO(10)$ singlets with $U(1)_{1,2,3}$ charges. The quantum 
numbers of the massless states from these two sectors are the 
same as those that are given in Ref. \cite{custodial}. 

The states from the sectors $b_j\oplus1+\alpha+2\gamma~(j=1,2,3)$ produce
the three light generations. The states from these sectors and their
decomposition under the entire gauge group are shown in table 2. 
The leptons are singlets of the color $SU(4)$ gauge group and 
the $U(1)_B$ symmetry, Eq. \ref{u1b} becomes the gauged leptophobic 
$U(1)$ symmetry. The remaining massless states and their quantum 
numbers are given in table 2. 

We observe that the leptophobia of the $U(1)_B$ symmetry is obtained from 
a combination of $U(1)_{B-L}$ plus the three flavor symmetries 
$U(1)_{r_{4,5,6}}$. The $Q_C$ charges of the leptons from each of the sectors
$b_{1,2,3}$ are canceled by their charges under the flavor symmetries
$U(1)_{r_{4,5,6}}$. Miraculously, the charges of the leptons under the 
flavor $U(1)$ symmetries are such that the cancelation occurs for 
all the leptons, in all the sectors. Thus, the leptophobic $U(1)$ 
symmetry is generation blind. 

The massless spectrum of the string model contains three anomalous 
$U(1)$ symmetries: Tr${U_1}=24$, Tr${U_2}=24$, Tr${U_3}=24$. 
Of the three anomalous $U(1)$s,  two can be rotated by
an orthogonal transformation. One combination remains anomalous and is 
uniquely given by: ${U_A}=k\sum_j [{Tr {U(1)_j}}]U(1)_j$, 
where $j$ runs over all the anomalous $U(1)$s. 
The ``anomalous'' $U(1)_A$ is broken by the Dine--Seiberg--Witten 
mechanism \cite{dsw} in which some states in the massless string 
spectrum obtain nonvanishing VEVs that cancel the anomalous $U(1)$ 
D--term equation.  
Thus, the leptophobic $U(1)_B$ symmetry is anomaly free under the entire 
string spectrum and can be left unbroken down to low energies. 

We now examine whether similar leptophobic symmetries can arise 
in similar superstring models. The model of Ref. \cite{custodial} 
differs from the model of table 1 by a change of a GSO projection 
coefficient. In this model, two massless gauge boson from the sector
$1+\alpha+2\gamma$ enhance one of the $U(1)$ combinations to 
$SU(2)_{\rm custodial}$. 
The full massless spectrum and symmetries is given in 
Ref. \cite{custodial}. However, this phase change does not modify 
the charges of the states from the sectors $b_{1,2,3}$ under the 
flavor symmetries $U(1)_{r_{4,5,6}}$. Thus, the same combination
of $U(1)_{C}$ plus the flavor symmetries $U(1)_{r_{4,5,6}}$
is a leptophobic $U(1)$ symmetry. In this model the color $SU(3)$ 
group is not enhanced. 

Next, we examine the massless spectrum of the model of Ref. \cite{EU}. 
The boundary condition basis vectors and the entire massless spectrum 
are given in Ref. \cite{EU}. The gauge boson from the NS sector are 
the same. The gauge group, however, is not enhanced. The sectors 
$b_{1,2,3}$ produce the three chiral generations, which are 
charged under the same flavor symmetries. The charges, however, 
under the flavor symmetries differ from the charges in table 1. 
For example, examining the charges of the states from the sector 
$b_1$, 
\begin{eqnarray}
&&({e_L^c}+{u_L^c})_{{1\over2},0,0,{1\over2},0,0}+\nonumber\\
&&({d_L^c}+{N_L^c})_{{1\over2},0,0,{-{1\over2}},0,0}+\nonumber\\
&&(L)_{{1\over2},0,0,{1\over2},0,0}+(Q)_{{1\over2},0,0,-{1\over2},0,0},
\label{eu1b1}
\end{eqnarray}
we observe that $e_L^c$ and $L$ have like--sign charges under 
$U_{r_4}$. Since they carry opposite sign charges under $U(1)_C$, 
$U_{r_4}$ cannot be used to cancel the $B-L$ charge for both of these
states. Since, they carry like--sign charges also under $U(1)_{r_1}$
a leptophobic $U(1)$ cannot be made from these $U(1)$ symmetries. 
It also ought to be mentioned that in this model
the flavor symmetries $U_{r_{4,5,6}}$ are anomalous. Therefore, their
combination with $U(1)_C$ is not anomaly free and must be broken. 
Thus, the existence of a universal leptophobic $U(1)$ in the previous
model is nontrivial. 

Next, we comment on the charges under the horizontal symmetries in the 
model of Ref. \cite{FNY}. This model contains as well similar horizontal 
flavor symmetries $U(1)_{r_{4,5,6}}$. Examining the charges of the chiral 
generations under these symmetries we observe that in this case
the combination $U(1)_C+3U(1)_4-3U(1)_5-3U(1)_6$ could serve as a 
leptophobic $U(1)$ symmetry. This combination cancels the $U(1)_{B-L}$
charge of the charged and doublet leptons from the sectors $b_{1,2,3}$. 
Interestingly, the charges of the right handed neutrinos do not vanish 
under this symmetry. However, in this case \cite{FNY} the combination 
$U(1)_5+U(1)_6$ is not anomaly free and therefore cannot be a good 
leptophobic $U(1)$ symmetry. By changing a GSO projection coefficient
we may change the sign of the charges under these symmetries, which will 
flip the sign of the combination. However, this is not likely to help 
in this model as it might also change the sign of the anomaly. 

One possible interpretation of the $R_b$ and $R_c$ anomalies at LEP 
is the existence of an additional $Z^\prime$ which is universally 
decoupled from leptons. 
In this paper we examined how such leptophobic $U(1)$ symmetries may 
arise from superstring derived models. We showed in a specific 
toy model that the $U(1)_{B-L}$ gauge symmetry can combine with 
the horizontal flavor symmetries to produce a leptophobic 
$U(1)$ symmetry. The leptophobic $U(1)$ combination is 
universal and anomaly free. The appearance of such a symmetry 
seems to be nontrivial. It would be of further interest to 
examine whether other combinations of the flavor symmetries 
can produce universal leptophobic symmetries. It
would be also be of interest to examine whether leptophobic 
$U(1)$ symmetries can arise in other classes of superstring 
derived models \cite{lykken,ibanez}. In the class of free 
fermionic models that we studied in this paper, the universality 
of the leptophobic $U(1)$ symmetry is closely related to 
the underlying $Z_2\times Z_2$ orbifold structure, which is 
exhibited in the NAHE set. Due to this underlying structure
each one of the chiral generations is charged with respect 
to an orthogonal set of flavor quantum numbers. This property
enabled a combination of the flavor $U(1)$ symmetries to 
cancel the $U(1)_{B-L}$ charge of the leptons from each sector
separately, thus creating the universal leptophobic symmetry. 
It is also interesting to note that 
the appearance of a leptophobic symmetry
may be correlated with proton stability 
in the models that we examined in some detail. 
In this class of models proton decay from states at the massless 
string level is forbidden \cite{ps}. 
Finally, if the LEP anomalies persist and the $Z^\prime$ 
interpretation is verified in future experiments, this 
discovery will indicate the existence of some structure
which is beyond the Standard Model and may be beyond 
the simple scenarios of unification. Thus, such a 
discovery may be the first strong experimental evidence in favor
of superstring unification, in which such additional symmetries
are abundant and well motivated. It is of course important 
to examine in specific models whether such a leptophobic 
$Z^\prime$, which is obtained from a superstring model, 
can define a realistic low energy scenario. 
Such work is in progress.


It is a pleasure to thank Claudio Coriano and Pierre Ramond
for valuable discussions. This work was supported in part by 
DOE Grant No.\ DE-FG-0586ER40272 and by CICYT under contract
AEN94-0936.

\bibliographystyle{unsrt}

\vfill
\eject

\textwidth=7.5in
\oddsidemargin=-18mm
\topmargin=-5mm
\renewcommand{\baselinestretch}{1.3}
\pagestyle{empty}
\begin{table}
\begin{eqnarray*}
&\begin{tabular}{|c|c|ccc|cccccccc|cccccccc|}
\hline
 ~ & $\psi^\mu$ & $\chi^{12}$ & $\chi^{34}$ & $\chi^{56}$ &
        $\overline{\psi}^{1} $ &
        $\overline{\psi}^{2} $ &
        $\overline{\psi}^{3} $ &
        $\overline{\psi}^{4} $ &
        $\overline{\psi}^{5} $ &
        $\overline{\eta}^1 $&
        $\overline{\eta}^2 $&
        $\overline{\eta}^3 $&
        $\overline{\phi}^{1} $ &
        $\overline{\phi}^{2} $ &
        $\overline{\phi}^{3} $ &
        $\overline{\phi}^{4} $ &
        $\overline{\phi}^{5} $ &
        $\overline{\phi}^{6} $ &
        $\overline{\phi}^{7} $ &
        $\overline{\phi}^{8} $ \\
\hline
      $\alpha$ & 0 & 0 & 0 & 0 & 1 & 1 & 1 & 0 & 0 & 0 & 0 & 0 &
                 1 & 1 & 1 & 1 & 0 & 0 & 0 & 0 \\ 
      $\beta$ & 0 & 0 & 0 & 0 & 1 & 1 & 1 & 0 & 0 & 0 & 0 & 0 &
                 1 & 1 & 1 & 1 & 0 & 0 & 0 & 0 \\ 
      $\gamma$ & 0 & 0 & 0 & 0 & ${1\over 2}$ & ${1\over 2}$ & ${1\over 2}$ &
                 ${1\over 2}$ & ${1\over 2}$ & ${1\over 2}$ & ${1\over 2}$ &
                 ${1\over 2}$ & ${1\over 2}$ & 0 & 1 & 1 & ${1\over 2}$ &
                 ${1\over 2}$ & ${1\over 2}$ & 0 \\
\hline
\end{tabular}
   \nonumber\\
&  ~ \nonumber\\
&  ~ \nonumber\\
&\begin{tabular}{|c|cccc|cccc|cccc|}
\hline
 ~&     $y^3y^6$ & $y^4\overline y^4$ & $y^5\overline y^5$ &
        $\overline y^3\overline y^6$ & $y^1\omega^6$ & $y^2\overline y^2$ & 
        $\omega^5\overline \omega^5$ & $\overline y^1\overline\omega^6$ &
        $\omega^1\omega^3$ & $\omega^2\overline\omega^2$ & 
        $\omega^4\overline\omega^4$ & $\overline\omega^1\overline\omega^3$ \\ 
\hline
       $\alpha$ & 1 & 1 & 1 & 0 & 1 & 1 & 1 & 0 & 1 & 1 & 1 & 0 \\  
       $\beta$ & 0 & 1 & 0 & 1 & 0 & 1 & 0 & 1 & 1 & 0 & 0 & 0 \\  
       $\gamma$ & 0 & 0 & 1 & 1 & 1 & 0 & 0 & 0 & 0 & 1 & 0 & 1 \\  
\hline
\end{tabular}
\label{model}
\end{eqnarray*}
\caption{A three generation $SU(4)\times SU(2)\times U(1)$ model. 
The choice of generalized GSO coefficients is: 
$c{b_j\choose \alpha,\beta,\gamma}=
-c{\alpha\choose 1}=
-c{\alpha\choose \beta}=
-c{\beta\choose 1}=
 c{\gamma\choose 1}=
-c{\gamma\choose \alpha,\beta}=-1~$ 
$(j=1,2,3)$, with the others
specified by modular invariance and space--time supersymmetry.}
\end{table}

\begin{table}
\begin{eqnarray*}
\begin{tabular}{|c|c|c|rrrrrrr|c|rr|}
\hline
   $F$ & SEC & $SU(4)_C\times SU(2)_L$&$Q_{C'}$ & $Q_L$ & $Q_1$ & 
   $Q_2$ & $Q_3$ & $Q_{4'}$ & $Q_{5'}$ & $SU(5)_H\times SU(3)_H$ &
   $Q_{6'}$ & $Q_{8''}$ \\
\hline
   $L_1$ & $b_1 \oplus$ & $(1,2)$&$-{3\over 2}$ & $0$ & ${1\over 2}$ &
   $0$ & $0$ & $-{1\over 2}$ & $-{1\over 2}$ & $(1,1)$ & $-{8\over 3}$ &
   $0$ \\
   $Q_1$ & $1+\alpha+2\gamma$&$(4,2)$&${1\over 2}$&$0$&${1\over 2}$ & 
   $0$ & $0$ & $-{1\over 2}$ & $-{1\over 2}$ & $(1,1)$ & $-{2\over 3}$ &
   $0$ \\
   $d_1$ &  & $(\overline 4,1)$&$-{1\over 2}$ & $1$ & ${1\over 2}$ & 
   $0$ & $0$ & ${1\over 2}$ & ${1\over 2}$ & $(1,1)$ & ${2\over 3}$ &
   $0$ \\
   $N_1$ &  & $(1,1)$&${3\over 2}$ & $-1$ & ${1\over 2}$ &
   $0$ & $0$ & ${1\over 2}$ & ${1\over 2}$ & $(1,1)$ & ${8\over 3}$ &
   $0$ \\
   $e_1$ &  & $(1,1)$&${3\over 2}$ & $1$ & ${1\over 2}$ & 
   $0$ & $0$ & ${1\over 2}$ & ${1\over 2}$ & $(1,1)$ & ${8\over 3}$ &
   $0$ \\
   $u_1$ &  & $(\overline 4,1)$&$-{1\over 2}$ & $-1$ & ${1\over 2}$ & 
   $0$ & $0$ & ${1\over 2}$ & ${1\over 2}$ & $(1,1)$ & ${2\over 3}$ &
   $0$ \\
\hline
   $L_2$ & $b_2 \oplus$ & $(1,2)$&$-{3\over 2}$ & $0$&0 & ${1\over 2}$ &
    $0$ & ${1\over 2}$ & $-{1\over 2}$ & $(1,1)$ & $-{8\over 3}$ &
   $0$ \\
   $Q_2$ & $1+\alpha+2\gamma$&$(4,2)$&${1\over 2}$&$0$&0&${1\over 2}$ & 
    $0$ & ${1\over 2}$ & $-{1\over 2}$ & $(1,1)$ & $-{2\over 3}$ &
   $0$ \\
   $d_2$ &  & $(\overline 4,1)$&$-{1\over 2}$ & $1$&0 & ${1\over 2}$ & 
    $0$ & $-{1\over 2}$ & ${1\over 2}$ & $(1,1)$ & ${2\over 3}$ &
   $0$ \\
   $N_2$ &  & $(1,1)$&${3\over 2}$ & $-1$ & 0 & ${1\over 2}$ &
    $0$ & $-{1\over 2}$ & ${1\over 2}$ & $(1,1)$ & ${8\over 3}$ &
   $0$ \\
   $e_2$ &  & $(1,1)$&${3\over 2}$ & $1$&0 & ${1\over 2}$ & 
    $0$ & $-{1\over 2}$ & ${1\over 2}$ & $(1,1)$ & ${8\over 3}$ &
   $0$ \\
   $u_2$ &  & $(\overline 4,1)$&$-{1\over 2}$ & $-1$&0 & ${1\over 2}$ & 
    $0$ & $-{1\over 2}$ & ${1\over 2}$ & $(1,1)$ & ${2\over 3}$ &
   $0$ \\
\hline
   $L_3$ & $b_3 \oplus$ & $(1,2)$&$-{3\over 2}$ & $0$&0&0 & ${1\over 2}$ &
   $0$ & ${1}$ & $(1,1)$ & $-{8\over 3}$ &
   $0$ \\
   $Q_3$ & $1+\alpha+2\gamma$&$(4,2)$&${1\over 2}$&$0$&0&0&${1\over 2}$ &
    $0$ & ${1}$ & $(1,1)$ & $-{2\over 3}$ &
   $0$ \\
   $d_3$ &  & $(\overline 4,1)$&$-{1\over 2}$ & $1$&0&0 & ${1\over 2}$ &
    $0$ & $-{1}$ & $(1,1)$ & ${2\over 3}$ &
   $0$ \\
   $N_3$ &  & $(1,1)$&${3\over 2}$ & $-1$&0&0 & ${1\over 2}$ &
    $0$ & $-{1}$ & $(1,1)$ & ${8\over 3}$ &
   $0$ \\
   $e_3$ &  & $(1,1)$&${3\over 2}$ & $1$&0&0 & ${1\over 2}$ &
    $0$ & $-{1}$ & $(1,1)$ & ${8\over 3}$ &
   $0$ \\
   $u_3$ &  & $(\overline 4,1)$&$-{1\over 2}$ & $-1$&0&0 & ${1\over 2}$ &
    $0$ & $-{1}$ & $(1,1)$ & ${2\over 3}$ &
   $0$ \\
\hline
\end{tabular}
\label{matter1}
\end{eqnarray*}
\caption{Three generations of massless states and their quantum numbers in the
model of Table 1.} 
\end{table}

\begin{table}
\begin{eqnarray*}
\begin{tabular}{|c|c|c|rrrrrrr|c|rr|}
\hline
   $F$ & SEC & $SU(4)_C\times SU(2)_L$&$Q_{C'}$ & $Q_L$ & $Q_1$ & 
   $Q_2$ & $Q_3$ & $Q_{4'}$ & $Q_{5'}$ & $SU(5)_H\times SU(3)_H$ &
   $Q_{6'}$ & $Q_{8''}$ \\
\hline
    $V_1$ & $b_1+2\gamma$ & $(1,1)$&$-{1\over 2}$ & $0$ & $0$ &
   ${1\over 2}$ & ${1\over 2}$ & ${1\over 2}$ & ${1\over 2}$ & 
   $(1,3)$ & ${8\over 3}$ & ${5\over 2}$ \\
    $\overline V_1$ & & $(1,1)$&${1\over 2}$ & $0$ & $0$ & 
   ${1\over 2}$ & ${1\over 2}$ & $-{1\over 2}$ & $-{1\over 2}$ & 
   $(1,{\bar3})$ & $-{8\over 3}$ & $-{5\over 2}$ \\
    $T_1$ &  & $(1,1)$&${1\over 2}$ & $0$ & $0$ &
   ${1\over 2}$ & ${1\over 2}$ & $-{1\over 2}$ & $-{1\over 2}$ & 
   $({5},1)$ & $-{8\over 3}$ & ${3\over 2}$ \\
    $\overline T_1$ &  & $(1,1)$&$-{1\over 2}$ & $0$ & $0$ & 
   ${1\over 2}$ & ${1\over 2}$ & ${1\over 2}$ & ${1\over 2}$ & 
   $({\bar5,}1)$ & ${8\over 3}$ & $-{3\over 2}$ \\
\hline
    $V_2$ & $b_2+2\gamma$ & $(1,1)$&$-{1\over 2}$ & $0$ & 
   ${1\over 2}$&0 & ${1\over 2}$ & $-{1\over 2}$ & ${1\over 2}$ & 
   $(1,3)$ & ${8\over 3}$ & ${5\over 2}$ \\
    $\overline V_2$ & & $(1,1)$&${1\over 2}$ & $0$ &  
   ${1\over 2}$&0 & ${1\over 2}$ & ${1\over 2}$ & $-{1\over 2}$ & 
   $(1,{\bar3})$ & $-{8\over 3}$ & $-{5\over 2}$ \\
    $T_2$ &  & $(1,1)$&${1\over 2}$ & $0$ & 
   ${1\over 2}$&0 & ${1\over 2}$ & ${1\over 2}$ & $-{1\over 2}$ & 
   $(5,1)$ & $-{8\over 3}$ & ${3\over 2}$ \\
    $\overline T_2$ &  & $(1,1)$&$-{1\over 2}$ & $0$ &  
   ${1\over 2}$&0 & ${1\over 2}$ & $-{1\over 2}$ & ${1\over 2}$ & 
   $({\bar5},1)$ & ${8\over 3}$ & $-{3\over 2}$ \\
\hline
    $V_3$ & $b_3+2\gamma$ & $(1,1)$&$-{1\over 2}$ & $0$ & 
   ${1\over 2}$ & ${1\over 2}$&0 & $0$ & $-1$ & 
   $(1,3)$ & ${8\over 3}$ & ${5\over 2}$ \\
    $\overline V_3$ & & $(1,1)$&${1\over 2}$ & $0$ &  
   ${1\over 2}$ & ${1\over 2}$&0 & $0$ & $1$ & 
   $(1,{\bar3})$ & $-{8\over 3}$ & $-{5\over 2}$ \\
    $T_3$ &  & $(1,1)$&${1\over 2}$ & $0$ & 
   ${1\over 2}$ & ${1\over 2}$&0 & $0$ & $1$ & 
   $(5,1)$ & $-{8\over 3}$ & ${3\over 2}$ \\
    $\overline T_3$ &  & $(1,1)$&$-{1\over 2}$ & $0$ &  
   ${1\over 2}$ & ${1\over 2}$&0 & $0$ & $-1$ & 
   $({\bar5},1)$ & ${8\over 3}$ & $-{3\over 2}$ \\
\hline
   $l_1$ & $b_2+b_3+$ & $(1,2)$&$-{3\over 4}$ &
    $-{1\over 2}$ & ${1\over 4}$ & $-{1\over 4}$ & $-{1\over 4}$ &
    $0$ & $0$ & $(1,1)$ & $0$ & $-{15\over 4}$ \\
   $\overline l_1$ &$\beta+\gamma+\xi$  & $(1,\overline 2)$&${3\over 4}$ &
    ${1\over 2}$ & $-{1\over 4}$ & ${1\over 4}$ & ${1\over 4}$ &
    $0$ & $0$ & $(1,1)$ & $0$ & ${15\over 4}$ \\
   $S_1$ &  & $(1,1)$&${3\over 4}$ &
    $-{1\over 2}$ & $-{3\over 4}$ & $-{1\over 4}$ & $-{1\over 4}$ &
    $0$ & $0$ & $(1,1)$ & $0$ & $-{15\over 4}$ \\
   $\overline S_1$ &  & $(1,1)$&$-{3\over 4}$ &
    ${1\over 2}$ & ${3\over 4}$ & ${1\over 4}$ & ${1\over 4}$ &
    $0$ & $0$ & $(1,1)$ & $0$ & ${15\over 4}$ \\
   $S_2$ &  & $(1,1)$&${3\over 4}$ &
    $-{1\over 2}$ & $-{1\over 4}$ & $-{3\over 4}$ & ${1\over 4}$ &
    $0$ & $0$ & $(1,1)$ & $0$ & $-{15\over 4}$ \\
   $\overline S_2$ &  & $(1,1)$&$-{3\over 4}$ &
    ${1\over 2}$ & ${1\over 4}$ & ${3\over 4}$ & $-{1\over 4}$ &
    $0$ & $0$ & $(1,1)$ & $0$ & ${15\over 4}$ \\
   $S_3$ &  & $(1,1)$&${3\over 4}$ &
    $-{1\over 2}$ & $-{1\over 4}$ & ${1\over 4}$ & $-{3\over 4}$ &
    $0$ & $0$ & $(1,1)$ & $0$ & $-{15\over 4}$ \\
   $\overline S_3$ &  & $(1,1)$&$-{3\over 4}$ &
    ${1\over 2}$ & ${1\over 4}$ & $-{1\over 4}$ & ${3\over 4}$ &
    $0$ & $0$ & $(1,1)$ & $0$ & ${15\over 4}$ \\
   $H_1$ &  & $(1,1)$&$-{3\over 4}$ &
    ${1\over 2}$ & ${1\over 4}$ & $-{1\over 4}$ & $-{1\over 4}$ &
    $0$ & $0$ & $(5,1)$ & $0$ & ${9\over 4}$ \\
   $\overline H_1$ &  & $(1,1)$&${3\over 4}$ &
    $-{1\over 2}$ & $-{1\over 4}$ & ${1\over 4}$ & ${1\over 4}$ &
    $0$ & $0$ & $({\bar5},1)$ & $0$ & $-{9\over 4}$ \\
\hline
   $l_2$ & $b_1+b_3+$ & $(1,2)$&$-{3\over 4}$ &
    $-{1\over 2}$ & ${1\over 4}$ & $-{1\over 4}$ & $-{1\over 4}$ &
    $0$ & $0$ & $(1,1)$ & $0$ & $-{15\over 4}$ \\
   $\overline l_2$ & $\alpha+\gamma+\xi$  & $(1,\overline 2)$&${3\over 4}$ &
    ${1\over 2}$ & $-{1\over 4}$ & ${1\over 4}$ & ${1\over 4}$ &
    $0$ & $0$ & $(1,1)$ & $0$ & ${15\over 4}$ \\
   $S_4$ &  & $(1,1)$&${3\over 4}$ &
    $-{1\over 2}$ & $-{3\over 4}$ & $-{1\over 4}$ & ${1\over 4}$ &
    $0$ & $0$ & $(1,1)$ & $0$ & $-{15\over 4}$ \\
   $\overline S_4$ &  & $(1,1)$&$-{3\over 4}$ &
    ${1\over 2}$ & ${3\over 4}$ & ${1\over 4}$ & $-{1\over 4}$ &
    $0$ & $0$ & $(1,1)$ & $0$ & ${15\over 4}$ \\
   $S_5$ &  & $(1,1)$&${3\over 4}$ &
    $-{1\over 2}$ & $-{1\over 4}$ & $-{3\over 4}$ & $-{1\over 4}$ &
    $0$ & $0$ & $(1,1)$ & $0$ & $-{15\over 4}$ \\
   $\overline S_5$ &  & $(1,1)$&$-{3\over 4}$ &
    ${1\over 2}$ & ${1\over 4}$ & ${3\over 4}$ & ${1\over 4}$ &
    $0$ & $0$ & $(1,1)$ & $0$ & ${15\over 4}$ \\
   $S_6$ &  & $(1,1)$&${3\over 4}$ &
    $-{1\over 2}$ & ${1\over 4}$ & $-{1\over 4}$ & $-{3\over 4}$ &
    $0$ & $0$ & $(1,1)$ & $0$ & $-{15\over 4}$ \\
   $\overline S_6$ &  & $(1,1)$&$-{3\over 4}$ &
    ${1\over 2}$ & $-{1\over 4}$ & ${1\over 4}$ & ${3\over 4}$ &
    $0$ & $0$ & $(1,1)$ & $0$ & ${15\over 4}$ \\
   $H_2$ &  & $(1,1)$&$-{3\over 4}$ &
    ${1\over 2}$ & $-{1\over 4}$ & ${1\over 4}$ & $-{1\over 4}$ &
    $0$ & $0$ & $(5,1)$ & $0$ & ${9\over 4}$ \\
   $\overline H_2$ &  & $(1,1)$&${3\over 4}$ &
    $-{1\over 2}$ & ${1\over 4}$ & $-{1\over 4}$ & ${1\over 4}$ &
    $0$ & $0$ & $({\bar5},1)$ & $0$ & $-{9\over 4}$ \\
\hline
\end{tabular}
\label{matter2}
\end{eqnarray*}
\caption{Extra massless states and their quantum numbers in the
model of Table 1.} 
\end{table}

\begin{table}
\begin{eqnarray*}
\begin{tabular}{|c|c|c|rrrrrrr|c|rr|}
\hline
   $F$ & SEC & $SU(4)_C\times SU(2)_L$&$Q_{C'}$ & $Q_L$ & $Q_1$ & 
   $Q_2$ & $Q_3$ & $Q_{4'}$ & $Q_{5'}$ & $SU(5)_H\times SU(3)_H$ &
   $Q_{6'}$ & $Q_{8''}$ \\
\hline
   $l_4$ & $1+b_1+$ & $(1,2)$&$-1$ &
    $0$ & $-{1\over 2}$ & 0  & 0 &
    $-{1\over 2}$ & $-{1\over 2}$ & $(1,1)$ & ${16\over 3}$ & $0$ \\
   $S_7$ &$\alpha+2\gamma$  & $(1,1)$&$1$ &
    $1$ & $-{1\over 2}$ & $0$ & $0$ &
    ${1\over 2}$ & ${1\over 2}$ & $(1,1)$ & $-{16\over 3}$ & $0$ \\
   $\overline S_7$ &  & $(1,1)$&$1$ &
    $-1$ & $-{1\over 2}$ & $0$ & $0$ &
    ${1\over 2}$ & ${1\over 2}$ & $(1,1)$ & $-{16\over 3}$ & $0$ \\
\hline
   $l_5$ & $1+b_2+$ & $(1,2)$&$-1$ &
    $0$ & 0 & $-{1\over 2}$  & 0 &
    ${1\over 2}$ & $-{1\over 2}$ & $(1,1)$ & ${16\over 3}$ & $0$ \\
   $S_8$ &$\alpha+2\gamma$  & $(1,1)$&$1$ &
    $1$ & 0 & $-{1\over 2}$ & $0$ &
    $-{1\over 2}$ & ${1\over 2}$ & $(1,1)$ & $-{16\over 3}$ & $0$ \\
   $\overline S_8$ &  & $(1,1)$&$1$ &
    $-1$ & 0 & $-{1\over 2}$ & $0$ &
    $-{1\over 2}$ & ${1\over 2}$ & $(1,1)$ & $-{16\over 3}$ & $0$ \\
\hline
   $l_6$ & $1+b_3+$ & $(1,2)$&$-1$ &
    $0$ & $0$ & $0$ & $-{1\over 2}$ &
    $0$ & $1$ & $(1,1)$ & ${16\over 3}$ & $0$ \\
   $S_9$ &$\alpha+2\gamma$  & $(1,1)$&$1$ &
    $1$ & $0$ & $0$ & $-{1\over 2}$ &
    $0$ & $-1$ & $(1,1)$ & $-{16\over 3}$ & $0$ \\
   $\overline S_9$ &  & $(1,1)$&$1$ &
    $-1$ & $0$ & $0$ & $-{1\over 2}$ &
    $0$ & $-1$ & $(1,1)$ & $-{16\over 3}$ & $0$ \\
\hline
   $S_{10}$ &$1+s+$  & $(1,1)$& $-2 $ &
    $0$ & $0$ & $0$ & $0$ &
    $-1$ & $-1$ & $(1,1)$ & $-{4\over 3}$ & $0$ \\
   $\overline S_{10}$ & $\alpha+2\gamma$  & $(1,1)$&$-2$ &
    $0$ & $0$ & $0$ & $0$ &
    $1$ & $1$ & $(1,1)$ & ${4\over 3}$ & $0$ \\
\hline
\end{tabular}
\label{matter3}
\end{eqnarray*}
\caption{{\it (Cont.)} Extra massless states and their quantum numbers in the
model of Table 1.} 
\end{table}

\end{document}